\documentclass[prd,twocolumn,lengthcheck,superscriptaddress,showpacs,amssymb,amsmath,amsfonts,aps,altaffilletter,nofootinbib]{revtex4}

\usepackage{color}
\usepackage{times}
\usepackage{graphicx}
\usepackage{fancyhdr}
\usepackage{float}
\usepackage{ulem}
\usepackage[raggedright]{subfigure}
\usepackage{acronym}
\usepackage{multirow}
\usepackage{array}

\makeatletter
\makeatletter
\def\@fnsymbol#1{\ifcase#1\or * \or  $+$ \or  \$ \or \#  \or \dag \or \ddag \or
$\mathsection$ \or $ \mathparagraph$ \or $\|$  \or \textordfeminine \or \textbul
let   
\or ** \or $++$ \or  \$\$ \or \#\#  \or \dag\dag \or \ddag\ddag \or
$\mathsection\mathsection$ \or $ \mathparagraph\mathparagraph$ \or $\|\|$  \or 
\textordfeminine\textordfeminine \or \textbullet \textbullet \or *** \or $+++$ 
\or  \$\$\$ \or \#\#  \or \dag\dag \or \ddag\ddag \or
$\mathsection \mathsection\mathsection$ \or $ \mathparagraph 
\mathparagraph\mathparagraph$ \or $\|\|\|$  \or 
\textordfeminine\textordfeminine\textordfeminine \or 
\textbullet\textbullet\textbullet \or \else \@ctrerr\fi}
\makeatother

\newcommand\fake[1]{\textcolor{red}{#1}}

\def\thercsid{\relax}
\def\rcsid#1{\def\next##1#1{\def\thercsid{##1}}\next}

\rcsid$Id: s5-2yr-12to18-lowcbc.tex,v 1.122 2009/10/05 10:41:23 mckechan Exp $

\renewcommand{\today}{\number\day\space\ifcase\month\or
  January\or February\or March\or April\or May\or June\or
  July\or August\or September\or October\or November\or December\fi
  \space\number\year}

\def\Msun{\ensuremath{M_{\odot}}}
\def\BNSul{\ensuremath{1.4 \times 10^{-2}}}
\def\BHNSul{\ensuremath{3.6 \times 10^{-3}}}
\def\BBHul{\ensuremath{7.3 \times 10^{-4}}}

\def\SBHNSul{\ensuremath{4.4 \times 10^{-3}}}
\def\SBBHul{\ensuremath{9.0 \times 10^{-4}}}

\def\BNStripleCumLum{\ensuremath{490}}
\def\BNSHoneLoneCumLum{\ensuremath{410}}
\def\BNSHtwoLoneCumLum{\ensuremath{110}}

\def\BNStripleCalErr{\ensuremath{23\%}}
\def\BNSHoneLoneCalErr{\ensuremath{23\%}}
\def\BNSHtwoLoneCalErr{\ensuremath{26\%}}

\def\BNStripleMonErr{\ensuremath{3\%}}
\def\BNSHoneLoneMonErr{\ensuremath{7\%}}
\def\BNSHtwoLoneMonErr{\ensuremath{10\%}}

\def\BNStripleWavErr{\ensuremath{31\%}}
\def\BNSHoneLoneWavErr{\ensuremath{32\%}}
\def\BNSHtwoLoneWavErr{\ensuremath{31\%}}

\def\BNStripleGDErr{\ensuremath{16\%}}
\def\BNSHoneLoneGDErr{\ensuremath{16\%}}
\def\BNSHtwoLoneGDErr{\ensuremath{3\%}}

\def\BNStripleGMErr{\ensuremath{19\%}}
\def\BNSHoneLoneGMErr{\ensuremath{19\%}}
\def\BNSHtwoLoneGMErr{\ensuremath{17\%}}
\begin{document}

\acrodef{BBH}{binary black holes}
\acrodef{BNS}{binary neutron stars}
\acrodef{BHNS}{black hole--neutron star binary}
\acrodef{PBH}{primordial black hole binaries}
\acrodef{SNR}{signal-to-noise ratio}
\acrodef{SPA}{stationary-phase approximation}
\acrodef{LIGO}{Laser Interferometer Gravitational-wave Observatory}
\acrodef{LHO}{LIGO Hanford Observatory}
\acrodef{LLO}{LIGO Livingston Observatory}
\acrodef{LSC}{LIGO Scientific Collaboration}
\acrodef{GRB}{gamma-ray bursts}
\acrodef{CBC}{compact binary coalescence}
\acrodef{GW}{gravitational wave}
\acrodef{ISCO}{innermost stable circular orbit}
\acrodef{FAR}{false alarm rate}
\acrodef{IFAR}{inverse false alarm rate}
\acrodef{CL}{confidence level}
\acrodef{PN}{Post-Newtonian}
\acrodef{DQ}{data quality}
\acrodef{S5YR1}{S5 first year search}
\acrodef{S5YR2A}{S5 second year, pre VSR1}
\acrodef{VSR1}{first Virgo science run}
\acrodef{IFO}{interferometer}

\title{Search for Gravitational Waves from Low Mass 
Compact Binary Coalescence in 186 Days of LIGO's Fifth Science Run\\
{\color{red}\large LIGO-P0900009-v11}
}

%\author{The LSC}
%\input{T0900008-v7}
%********************* cut here mark*******************
%NOTE ADD altafillletter to \documentclass if lettered affiliation footnotes are desired
% EXAMPLE:
%
%\documentclass[prd,superscriptaddress,showpacs,amssymb,amsmath,amsfonts,aps,altaffilletter]{revtex4}
%
%%%%%%%%%%%% Institutes Number and definitions %%%%%%%%%%%
%
% This list was generated on 2 Apr 2009 A4 DCC LIGO - T0900008-v6
%
% to this C.D.Capano{\SR} and G.Ely{\CL} and A.P.Lundgren{\SR}
% have been added

\newcommand*{\AG}{Albert-Einstein-Institut, Max-Planck-Institut f\"{u}r Gravitationsphysik, D-14476 Golm, Germany}
\affiliation{\AG}
\newcommand*{\AH}{Albert-Einstein-Institut, Max-Planck-Institut f\"{u}r Gravitationsphysik, D-30167 Hannover, Germany}
\affiliation{\AH}
\newcommand*{\AU}{Andrews University, Berrien Springs, MI 49104 USA}
\affiliation{\AU}
\newcommand*{\AN}{Australian National University, Canberra, 0200, Australia}
\affiliation{\AN}
\newcommand*{\CH}{California Institute of Technology, Pasadena, CA  91125, USA}
\affiliation{\CH}
\newcommand*{\CA}{Caltech-CaRT, Pasadena, CA  91125, USA}
\affiliation{\CA}
\newcommand*{\CU}{Cardiff University, Cardiff, CF24 3AA, United Kingdom}
\affiliation{\CU}
\newcommand*{\CL}{Carleton College, Northfield, MN  55057, USA}
\affiliation{\CL}
\newcommand*{\CS}{Charles Sturt University, Wagga Wagga, NSW 2678, Australia}
\affiliation{\CS}
\newcommand*{\CO}{Columbia University, New York, NY  10027, USA}
\affiliation{\CO}
\newcommand*{\ER}{Embry-Riddle Aeronautical University, Prescott, AZ   86301 USA}
\affiliation{\ER}
\newcommand*{\EU}{E\"{o}tv\"{o}s University, ELTE 1053 Budapest, Hungary}
\affiliation{\EU}
\newcommand*{\HC}{Hobart and William Smith Colleges, Geneva, NY  14456, USA}
\affiliation{\HC}
\newcommand*{\IA}{Institute of Applied Physics, Nizhny Novgorod, 603950, Russia}
\affiliation{\IA}
\newcommand*{\IU}{Inter-University Centre for Astronomy  and Astrophysics, Pune - 411007, India}
\affiliation{\IU}
\newcommand*{\HU}{Leibniz Universit\"{a}t Hannover, D-30167 Hannover, Germany}
\affiliation{\HU}
\newcommand*{\CT}{LIGO - California Institute of Technology, Pasadena, CA  91125, USA}
\affiliation{\CT}
\newcommand*{\LO}{LIGO - Hanford Observatory, Richland, WA  99352, USA}
\affiliation{\LO}
\newcommand*{\LV}{LIGO - Livingston Observatory, Livingston, LA  70754, USA}
\affiliation{\LV}
\newcommand*{\LM}{LIGO - Massachusetts Institute of Technology, Cambridge, MA 02139, USA}
\affiliation{\LM}
\newcommand*{\LU}{Louisiana State University, Baton Rouge, LA  70803, USA}
\affiliation{\LU}
\newcommand*{\LE}{Louisiana Tech University, Ruston, LA  71272, USA}
\affiliation{\LE}
\newcommand*{\LL}{Loyola University, New Orleans, LA 70118, USA}
\affiliation{\LL}
\newcommand*{\MT}{Montana State University, Bozeman, MT 59717, USA}
\affiliation{\MT}
\newcommand*{\MS}{Moscow State University, Moscow, 119992, Russia}
\affiliation{\MS}
\newcommand*{\ND}{NASA/Goddard Space Flight Center, Greenbelt, MD  20771, USA}
\affiliation{\ND}
\newcommand*{\NA}{National Astronomical Observatory of Japan, Tokyo  181-8588, Japan}
\affiliation{\NA}
\newcommand*{\NO}{Northwestern University, Evanston, IL  60208, USA}
\affiliation{\NO}
\newcommand*{\RI}{Rochester Institute of Technology, Rochester, NY  14623, USA}
\affiliation{\RI}
\newcommand*{\RA}{Rutherford Appleton Laboratory, HSIC, Chilton, Didcot, Oxon OX11 0QX United Kingdom}
\affiliation{\RA}
\newcommand*{\SJ}{San Jose State University, San Jose, CA 95192, USA}
\affiliation{\SJ}
\newcommand*{\SM}{Sonoma State University, Rohnert Park, CA 94928, USA}
\affiliation{\SM}
\newcommand*{\SE}{Southeastern Louisiana University, Hammond, LA  70402, USA}
\affiliation{\SE}
\newcommand*{\SO}{Southern University and A\&M College, Baton Rouge, LA  70813, USA}
\affiliation{\SO}
\newcommand*{\SA}{Stanford University, Stanford, CA  94305, USA}
\affiliation{\SA}
\newcommand*{\SR}{Syracuse University, Syracuse, NY  13244, USA}
\affiliation{\SR}
\newcommand*{\PU}{The Pennsylvania State University, University Park, PA  16802, USA}
\affiliation{\PU}
\newcommand*{\UM}{The University of Melbourne, Parkville VIC 3010, Australia}
\affiliation{\UM}
\newcommand*{\MI}{The University of Mississippi, University, MS 38677, USA}
\affiliation{\MI}
\newcommand*{\SF}{The University of Sheffield, Sheffield S10 2TN, United Kingdom}
\affiliation{\SF}
\newcommand*{\TA}{The University of Texas at Austin, Austin, TX 78712, USA}
\affiliation{\TA}
\newcommand*{\TC}{The University of Texas at Brownsville and Texas Southmost College, Brownsville, TX  78520, USA}
\affiliation{\TC}
\newcommand*{\TR}{Trinity University, San Antonio, TX  78212, USA}
\affiliation{\TR}
\newcommand*{\BB}{Universitat de les Illes Balears, E-07122 Palma de Mallorca, Spain}
\affiliation{\BB}
\newcommand*{\UA}{University of Adelaide, Adelaide, SA 5005, Australia}
\affiliation{\UA}
\newcommand*{\BR}{University of Birmingham, Birmingham, B15 2TT, United Kingdom}
\affiliation{\BR}
\newcommand*{\FA}{University of Florida, Gainesville, FL  32611, USA}
\affiliation{\FA}
\newcommand*{\GU}{University of Glasgow, Glasgow, G12 8QQ, United Kingdom}
\affiliation{\GU}
\newcommand*{\MD}{University of Maryland, College Park, MD 20742 USA}
\affiliation{\MD}
\newcommand*{\AM}{University of Massachusetts - Amherst, Amherst, MA 01003, USA}
\affiliation{\AM}
\newcommand*{\MU}{University of Michigan, Ann Arbor, MI  48109, USA}
\affiliation{\MU}
\newcommand*{\MN}{University of Minnesota, Minneapolis, MN 55455, USA}
\affiliation{\MN}
\newcommand*{\OU}{University of Oregon, Eugene, OR  97403, USA}
\affiliation{\OU}
\newcommand*{\RO}{University of Rochester, Rochester, NY  14627, USA}
\affiliation{\RO}
\newcommand*{\SL}{University of Salerno, 84084 Fisciano (Salerno), Italy}
\affiliation{\SL}
\newcommand*{\SN}{University of Sannio at Benevento, I-82100 Benevento, Italy}
\affiliation{\SN}
\newcommand*{\SH}{University of Southampton, Southampton, SO17 1BJ, United Kingdom}
\affiliation{\SH}
\newcommand*{\SC}{University of Strathclyde, Glasgow, G1 1XQ, United Kingdom}
\affiliation{\SC}
\newcommand*{\WA}{University of Western Australia, Crawley, WA 6009, Australia}
\affiliation{\WA}
\newcommand*{\UW}{University of Wisconsin-Milwaukee, Milwaukee, WI  53201, USA}
\affiliation{\UW}
\newcommand*{\WU}{Washington State University, Pullman, WA 99164, USA}
\affiliation{\WU}

\author{}    \affiliation{\GU}    
\author{B.~P.~Abbott}    \affiliation{\CT}    
\author{R.~Abbott}    \affiliation{\CT}    
\author{R.~Adhikari}    \affiliation{\CT}    
\author{P.~Ajith}    \affiliation{\AH}    
\author{B.~Allen}    \affiliation{\AH}  \affiliation{\UW}  
\author{G.~Allen}    \affiliation{\SA}    
\author{R.~S.~Amin}    \affiliation{\LU}    
\author{S.~B.~Anderson}    \affiliation{\CT}    
\author{W.~G.~Anderson}    \affiliation{\UW}    
% ...
 \collaboration{The LIGO Scientific Collaboration, http://www.ligo.org}
 \noaffiliation
%
%
%********************* cut here mark*******************

%\date[\relax]{ RCS \thercsid; compiled \today }
\fake{}

\begin{abstract}

We report on a search for gravitational waves from coalescing compact binaries,
of total mass between $2$ and $35~\Msun$, using LIGO observations between 
November 14, 2006 and May 18, 2007. No gravitational-wave signals were
detected. We report upper limits on the rate of compact binary coalescence 
as a function of total mass. The LIGO cumulative 90\%-confidence rate upper
limits of the binary coalescence of neutron stars, black holes and
 black hole-neutron star systems are \BNSul, \BBHul and
\BHNSul $\textrm{ yr}^{-1} \mathrm{L}_{10}^{-1}$ respectively, where
$\mathrm{L}_{10}$ is $10^{10}$ times the blue solar luminosity.

\end{abstract}

\maketitle

%%%%%%%%%%%%%%%%%%%%%%%%%%%%%%%%%%%%%%
%\paragraph*{Introduction}\label{sec:overview}

In November 2005 the three first-generation detectors of the \ac{LIGO} reached
design sensitivity and began a two-year period of observations (known as the
fifth science run, or S5) which concluded in October 
2007~\cite{abbott:2007kva}.  One of the most promising sources of 
gravitational-waves for LIGO is a \ac{CBC}; the inspiral and merger of 
\ac{BNS}, \ac{BBH}, or a \ac{BHNS}~\cite{LIGOS1iul,LIGOS2iul,LIGOS2macho,
LIGOS2bbh,LIGOS3S4all,Collaboration:2009tt}. These systems spiral together as
they emit energy in the form of gravitational waves, finally merging to form a 
single object, which then settles down to equilibrium. Ground-based 
gravitational-wave detectors are most sensitive to waves with frequencies 
between $\sim 40$ and $1000$~Hz, corresponding to the late stages of inspiral 
and merger. In this paper we report the results of search for 
gravitational-waves from binaries with total mass between $2$ and $35~\Msun$ 
and a minimum component mass of $1~\Msun$ in LIGO observations between November 
14, 2006 and May 18, 2007. The results of a search for these systems in data 
taken from November 4, 2005 to November 14, 2006 were reported in 
Ref.~\cite{Collaboration:2009tt}. From May--October 2007, the Virgo 
gravitational-wave detector operated in coincidence with the \ac{LIGO} 
detectors~\cite{0264-9381-23-19-S01} and the \ac{LIGO} data from that period 
are being analyzed together with the Virgo data. The joint analysis requires 
significant modifications to our analysis pipeline: therefore results of that 
search will be reported in a subsequent publication. In contrast, the results 
presented here were obtained with substantially the same analysis pipeline used 
in Ref.~\cite{Collaboration:2009tt}.  

No gravitational-wave signals were observed during this search and so we
report upper limits on CBC rates using the upper limits of 
Ref.~\cite{Collaboration:2009tt} as prior rate distributions. We summarize the 
analysis procedure and we present the search results and upper limits on CBC 
rates derived from \ac{LIGO} observations in the period November 4, 2005 to May
18, 2007.

%%%%%%%%%%%%%%%%%%%%%%%%%%%%%%%%%%%%%%%
\paragraph*{ThedData analysis pipeline:}
\label{sec:pipeline}

The data-analysis pipeline used in this search is fundamentally the same as
that of Ref.~\cite{Collaboration:2009tt}, thus here we only describe the major 
components and highlight differences to the previous search, referring to 
Refs.~\cite{LIGOS3S4all,Collaboration:2009tt} for details. The most substantial
change in this analysis is a modification to the way in which the significance
of candidate events is compared to instrumental noise background. In previous
searches, the noise background was computed \emph{using the entire observation
period} by introducing an artificial time shift between data recorded at the
two LIGO observatories. The observation period is split into six four-week 
segments and one 18 day segment (referred to as ``months'') and the 
instrumental background is measured \emph{independently in each month}, as the
detector behavior varied over the course of the S5 run. Candidate triggers are
therefore compared to a background that better reflects the instrumental 
behavior at the time of the candidate.  Each month was searched independently 
for gravitational-wave candidates and in the absence of detections, the results
from the months are combined (together with the results from
Ref.~\cite{Collaboration:2009tt}) to set an upper limit on the CBC rate.

We search for gravitational-wave signals when at least two of the \ac{LIGO}
detectors were operational.  This comprised a total of $0.28$~yr when all three
detectors (the 4 and 2~km Hanford detectors, denoted H1 and H2, respectively, 
and the 4~km Livingston detector, denoted L1) were operational 
(H1H2L1 coincident data), $0.10$~yr of H1H2 coincident data, $0.02$~yr of H1L1 
coincident data, and $0.01$~yr of H2L1 coincident data.  Noise correlations 
between the colocated H1 and H2 detectors cause our method of estimating the 
instrumental background using time-shifted data to fail, and so we do not 
search data when only the H1H2 detectors are operating. Approximately $10\%$ of
data is designated \textit{playground} and used for tuning our search pipeline.

\ac{PN} theory provides accurate models of the inspiral waveform predicted by 
general relativity up to the
\ac{ISCO}~\cite{Blanchet:1996pi,Droz:1999qx,Blanchet:2002av,%
Buonanno:2006ui,Boyle:2007ft,Hannam:2007ik,%
pan:024014,Boyle:2009dg}. The frequency of the waveform from low mass binaries 
targeted in this search sweeps across the sensitive band of the LIGO detectors.
Therefore, we search for signals from our target sources by match filtering 
the data with \ac{PN} templates terminated at \ac{ISCO}. This method is 
suboptimal if a true signal differs from our template family due to unforeseen
physical effects. Matter effects in BNS and BHNS are not included in our 
templates, but are expected to be important only at higher frequencies~\cite{Shibata:2009cn,Kiuchi:2009jt}. We construct template banks~\cite{hexabank} of 
restricted second order \ac{PN} waveforms in the frequency
domain~\cite{thorne.k:1987,SathyaDhurandhar:1991,Droz:1999qx} such that no
more than $3\%$ of the \ac{SNR} is lost due to the discreteness of the 
bank~\cite{Owen:1998dk}. A ``trigger'' is generated if the matched-filter 
\ac{SNR} of the strain data filtered against the template exceeds a
threshold of $5.5$~\cite{Allen:2005fk}.  We demand that triggers are coincident 
in time of arrival and mass~\cite{Robinson:2008} in at least two of the 
three \ac{LIGO} detectors. When all three detector are operating we can obtain 
(in principle) four possible types of coincidence: H1H2L1 triple coincident 
triggers and three different double coincident types: H1H2, H1L1 and H2L1. 
We discard H1H2 double coincident triggers, due to the problems estimating the 
background for these triggers and discard H2L1 triggers when the H1 detector is
operating nominally (since the 4~km H1 detector is more sensitive than the 2~km 
H2 detector).
Coincident triggers are subjected to consistency checks using signal-based
vetoes~\cite{LIGOS3S4Tuning,Allen:2004,Rodriguez:2007}. Times of poor detector
data quality are flagged using environmental and auxiliary data;
triggers from these times are also vetoed~\cite{Collaboration:2009tt}. We
construct two categories of data-quality vetoes depending on the severity of
the instrumental artifact being flagged. In our primary search and upper limit
computation we veto coincident triggers that fall in times from either
category. We also consider detection candidates in data with only the most
severe category applied in case a loud signal is present that may
otherwise be vetoed.   Surviving triggers are clustered in time and ranked by
an effective \ac{SNR} statistic, which is computed from the trigger's
matched-filter \ac{SNR} and the value of the $\chi^2$ signal-based veto for
that trigger~\cite{LIGOS3S4all}.  After discarding playground data and times
in both veto categories, a total of $0.21$~yr of triple coincident data (H1H2L1)
,$0.02$~yr of H1L1 coincident data, and $0.01$~yr of H2L1 coincident data
remain. In the absence of a detection, these data are used to compute upper
limits on the \ac{CBC} rate.

\begin{figure}
\includegraphics[width=3in]{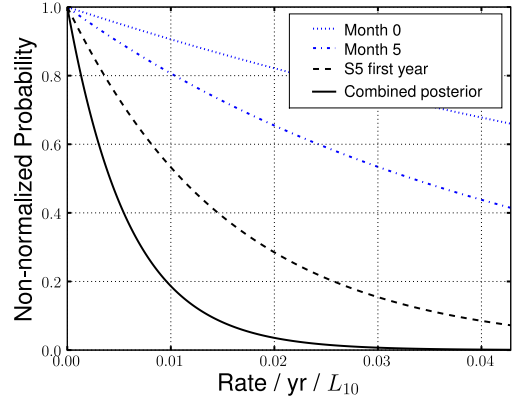}
\caption{The posterior distribution for the rate of \ac{BNS} coalescences. The
dashed black curve shows the rate computed in Ref.~\cite{Collaboration:2009tt}.
The solid black curve shows the result of this search using the previous
analysis as a prior. The figure also shows the rate distributions for two of
the  individual months computed using a uniform prior. The improvement from
month 0 to month 5 is due to increasing detector sensitivity during this
search.  }  
   \label{fig:ul}
\end{figure}

The rate of instrumental noise artifacts is measured by ime-shifting data from 
the Livingston and Hanford observatories (H1 and H2 data are kept fixed with 
respect to each other). The data are offset by more than the light-travel time 
between observatories, thus triggers which survive the pipeline are due to 
noise alone. We performed 100 such time-shifts to obtain a good estimate of 
the noise background in our search. \ac{CBC} signals of higher-mass contain 
fewer gravitational-wave cycles in the sensitive band of our detectors: 
our signal-based vetoes are not as powerful. High-mass templates are therefore 
more sensitive to nonstationary noise transients and hence our \ac{FAR} for 
these system is larger. In order to account for this mass-dependent behavior 
we compute the background for three different mass regions and compare 
foreground and background within each of these ranges. Specifically, in each
region we count the number of background triggers with effective
\ac{SNR} greater than or equal to a given foreground trigger; dividing
this number by the amount of background time analyzed gives us the
\ac{FAR} for that trigger. This allows us to define a single detection
statistic for every trigger in each of the mass categories.  The
\ac{FAR} can then be directly compared to obtain a ranking of the
significance of the triggers, regardless of their
mass~\cite{Collaboration:2009tt}. 

%%%%%%%%%%%%%%%%%%%%%%%%%%%%%%%%%%%%%%%
\paragraph*{Search results:}
\label{sec:results}

The seven months of data were analyzed separately using the procedure
described above. No gravitational-wave candidates were observed with a
\ac{FAR} significantly above those expected from the noise background.  The
loudest trigger in this search was a triple coincident trigger with a FAR of
$6$ per year. This is consistent with the expected background, since we
searched $0.21$~yr of data. The second and third loudest triggers had FAR
values of $10$ and $11$ per year respectively. Although we did not have any
detection candidates, we exercised our follow-up procedures by
examining any triggers with a \ac{FAR} of less than $50$ per year. 
This exercise prepares us for future detections and often identifies areas
where our search pipeline can be improved to exclude noise transients.

In the absence of detection candidates, we use our observations to set an
upper limit on the CBC rate. We follow the procedure described
in~\cite{Fairhurst:2007qj,loudestGWDAW03,Biswas:2007ni} 
and use the results reported in
Ref.~\cite{Collaboration:2009tt} as prior information on the rates.
We present five different classes of upper limits.  The
first three limits are placed on binaries of neutron stars and/or black
holes assuming canonical mass distributions for \ac{BNS} $[m_1 = m_2 =
(1.35\pm 0.04)~\Msun]$, \ac{BBH} $[m_1 = m_2 = (5\pm 1)~\Msun]$, and
\ac{BHNS} $[m_1 = (5\pm 1)~\Msun,~m_2 = (1.35\pm 0.04)~\Msun]$ systems.
We also present upper limits as a function of the total mass of the
binary and, for \ac{BHNS} binaries, as a function of the black hole
mass. We combine the results from each of the seven
months, along with the prior results from the first year analysis, in a
Bayesian manner, using the same procedure as described
in~\cite{Collaboration:2009tt}.

We first calculate upper limits on \ac{BNS}, \ac{BBH} and \ac{BHNS}
systems assuming the objects have no spin, and summarize the results
Tables \ref{tab:bns} and \ref{tab:ul}.
The rate of binary coalescences in a galaxy is expected to
be proportional to the blue light luminosity of the
galaxy~\cite{LIGOS3S4Galaxies}.  Therefore, we place limits on the rate
per $\mathrm{L}_{10}$ per year, where $\mathrm{L}_{10}$ is $10^{10}$
times the blue solar luminosity (the Milky Way contains $\sim 1.7
\mathrm{L}_{10}$~\cite{Kalogera:2000dz}).  To calculate the search
sensitivity, the analysis was repeated numerous times adding simulated
signals with a range of masses, distance and other astrophysical
parameters to the data. Table \ref{tab:ul} shows the sensitivity of 
the LIGO detectors to coalescing binaries quoted in terms 
of the horizon distance i.e., the distance at which an optimally oriented 
and located binary would produce an \ac{SNR} of 8.  
There are a number of uncertainties which affect the upper limit
calculation, including Monte Carlo statistics, detector calibration,
distances and luminosities of galaxies listed in the galaxy
catalog~\cite{LIGOS3S4Galaxies} and differences between the \ac{PN}
templates used to evaluate efficiency of the search and the actual
waveforms.  The effect of these errors on the cumulative luminosity are
summarized for the \ac{BNS} search in Table~\ref{tab:bns}.  We
marginalize over all of the uncertainties~\cite{Fairhurst:2007qj}
to obtain a posterior distribution on the rate of
binary coalescences.  

In Fig.~\ref{fig:ul}, we show the derived distribution of the rate of
\ac{BNS} coalescences. The distribution is peaked at zero rate
because there are no detection candidates.  We include the distribution for
all searches previous to this one (which is our prior).  In addition, we
present the result that would be obtained from each month, were it
analyzed independently of the others and of the previous searches.  This
provides an illustration of the amount that each month contributes to
the final upper limit result and demonstrates the improvement in
sensitivity of the detectors during the search.  The upper limit is
finally obtained by integrating the distribution from zero to
$\mathcal{R}_{90\%}$ so that $90\%$ of the probability is contained in the
interval.  The results obtained in this way are
%
%\begin{eqnarray}
$\mathcal{R}_{90\%,{\rm BNS}} = \BNSul\,
\textrm{yr}^{-1}\mathrm{L_{10}}^{-1} \, ,
\mathcal{R}_{90\%,{\rm BBH}} = \BBHul\,
\textrm{yr}^{-1}\mathrm{L_{10}}^{-1} \, , \text{ and }
\mathcal{R}_{90\%,{\rm BHNS}} =  \BHNSul\,
\textrm{yr}^{-1}\mathrm{L_{10}}^{-1} \, .$
%\end{eqnarray}

\begin{figure}[ht] 
  \includegraphics[width=3in]{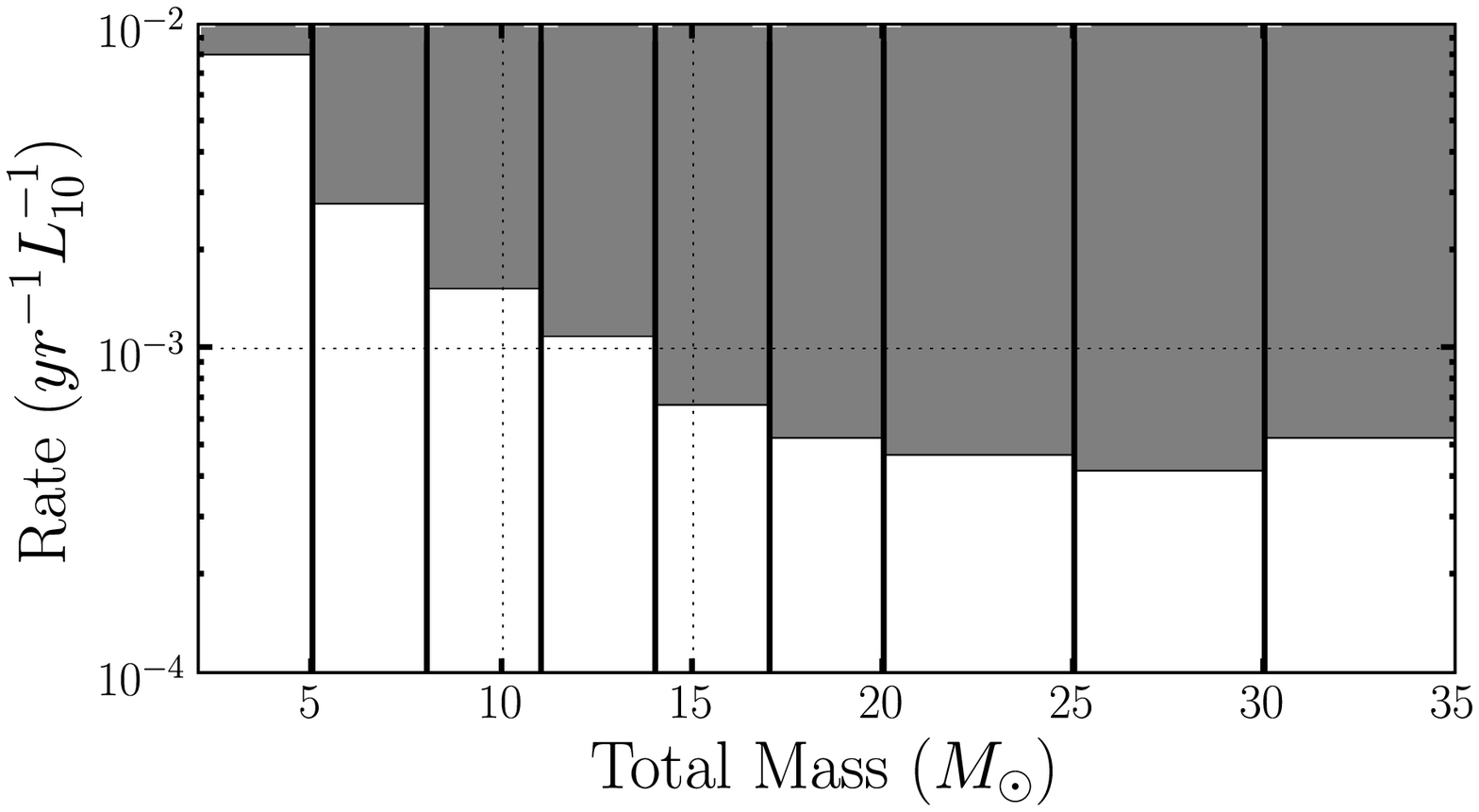}\vspace*{0.15cm}
  \includegraphics[width=3in]{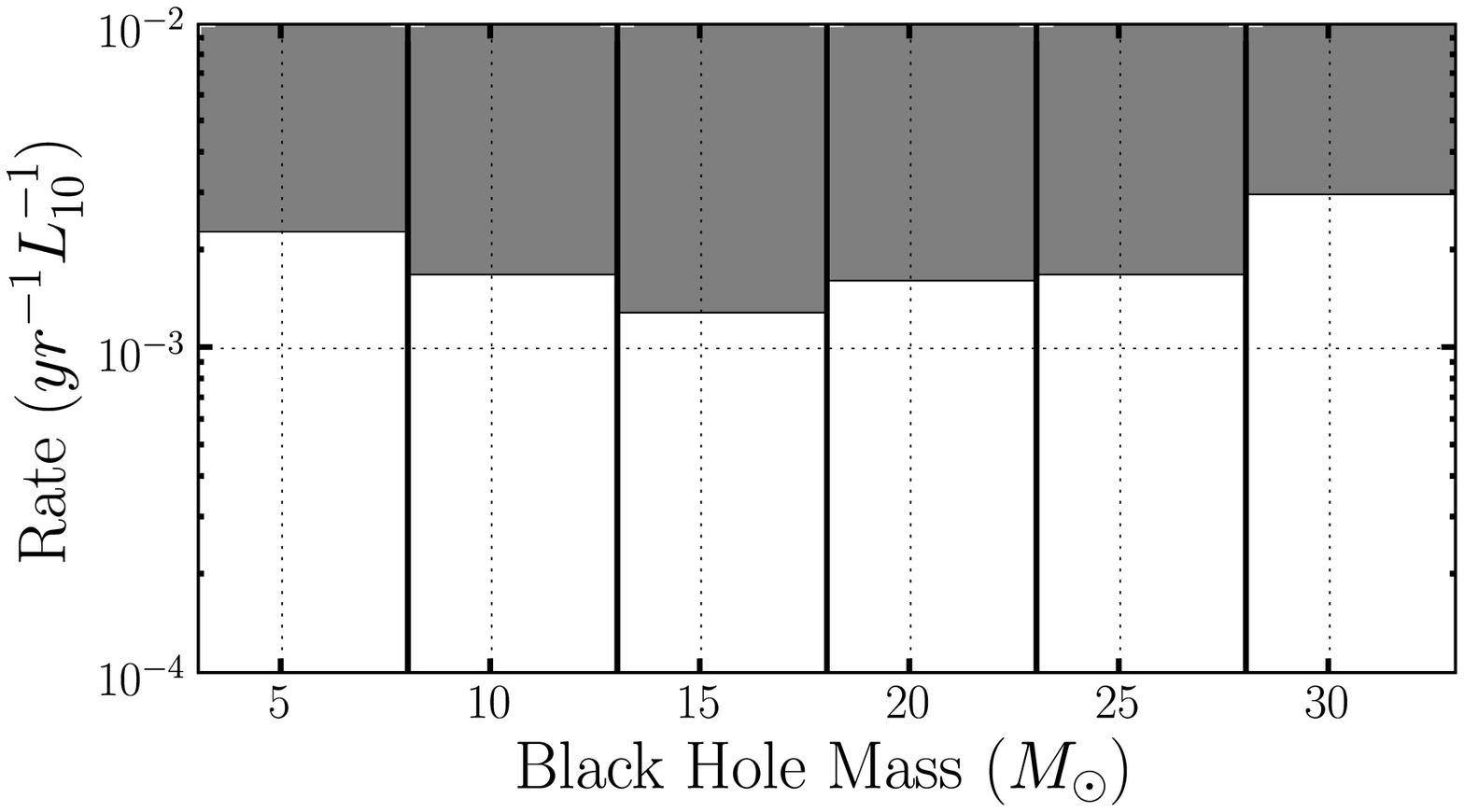}
  \caption{The marginalized 90\% rate upper limits as a function of mass.  The
upper plot shows limits for \ac{BBH} systems as a function of the
total mass of the system.  The lower plot shows limits for \ac{BHNS}
systems as a function of the black hole mass, assuming a fixed
neutron star mass of $1.35 M_{\odot}$. Here the upper limits are 
calculated using only H1H2L1 data since the relatively small amount
of H1L1 and H2L1 data makes it difficult to 
evaluate the cumulative luminosity in the individual mass bins.} 
  \label{fig:ulmass}
\end{figure}

Additionally we calculate the upper limit for \ac{BBH} systems
as a function of the total mass of the binary, assuming a uniform
distribution of the component masses.  For \ac{BHNS} systems, we
construct an upper limit as a function of the black hole mass, assuming
a fixed neutron star mass of $m_{\mathrm{NS}} = 1.35
\Msun$.  These upper limits are shown in Fig~\ref{fig:ulmass}.

\begin{table}[t]
\center
\begin{tabular}{c | c | c | c}
\hline \hline
\multicolumn{1}{m{5cm}|}{\centering Coincidence time} & H1H2L1 & H1L1 & H2L1 \\
\hline
\multicolumn{1}{m{5cm}|}{\centering Observation time (yr)} & 0.21 & 0.02 & 0.01 \\
\hline
\multicolumn{1}{m{5cm}|}{\centering Cumulative luminosity $\left({L_{10}}\right)$} & $\BNStripleCumLum$ & $\BNSHoneLoneCumLum$ & $\BNSHtwoLoneCumLum$ \\
\hline
\multicolumn{1}{m{5cm}|}{\centering Calibration error} & $\BNStripleCalErr$ & $\BNSHoneLoneCalErr$ & $\BNSHtwoLoneCalErr$ \\
\hline
\multicolumn{1}{m{5cm}|}{\centering Monte Carlo error} & $\BNStripleMonErr$ & $\BNSHoneLoneMonErr$ & $\BNSHtwoLoneMonErr$ \\
\hline
\multicolumn{1}{m{5cm}|}{\centering Waveform error} & $\BNStripleWavErr$ & $\BNSHoneLoneWavErr$ & $\BNSHtwoLoneWavErr$ \\
\hline
\multicolumn{1}{m{5cm}|}{\centering Galaxy distance error} & $\BNStripleGDErr$ & $\BNSHoneLoneGDErr$ & $\BNSHtwoLoneGDErr$ \\
\hline
\multicolumn{1}{m{5cm}|}{\centering Galaxy magnitude error} & $\BNStripleGMErr$ & $\BNSHoneLoneGMErr$ & $\BNSHtwoLoneGMErr$ \\
\hline
\hline
\end{tabular}
\caption{Detailed results from the \ac{BNS} search.  The observation
time is the time used in the upper limit analysis.  The cumulative
luminosity is the luminosity to which the search is sensitive above the
loudest event for each coincidence time.  The errors in this table are
listed as one-sigma logarithmic error bars (expressed as percentages) in
luminosity associated with each source error.}
\label{tab:bns}
\end{table}

\begin{table}[t]
\center
\begin{tabular}{c | c | c | c}
\hline \hline
\multicolumn{1}{m{3cm}|}{\centering Component masses $\left(M_{\odot}\right)$} & 1.35/1.35 & 5.0/5.0 & 5.0/1.35 \\
\hline
\multicolumn{1}{m{3cm}|}{\centering $D_{\rm horizon}$ $\left({\rm Mpc}\right)$} & $\sim 30$ & $\sim 100$ & $\sim 60$ \\
\hline
\multicolumn{1}{m{3cm}|}{\centering Cumulative lminosity $\left({L_{10}}\right)$} & 490 & 11000 & 2100 \\
\hline
\multicolumn{1}{m{3cm}|}{\centering Nonspinning upper limit $\left({{\rm yr}^{-1} L_{10}^{-1}}\right)$} & \BNSul & \BBHul & \BHNSul \\
\hline
\multicolumn{1}{m{3cm}|}{\centering Spinning upper limit $\left({{\rm yr}^{-1} L_{10}^{-1}}\right)$} & ... & \SBBHul & \SBHNSul \\
\hline
\hline
\end{tabular}
\caption{Overview of results from \ac{BNS}, \ac{BBH} and \ac{BHNS}
searches.  $D_{\rm horizon}$ is the horizon distance 
averaged over the time of the search.  The cumulative luminosity is the
luminosity to which the search is sensitive above the loudest event for
times when all three \ac{LIGO} detectors were operational.  The first
set of upper limits are those obtained for binaries with nonspinning
components.  The second set of upper limits are produced using black
holes with a spin uniformly distributed between zero and the maximal
value of $G m^{2}/c$.}
\label{tab:ul}
\end{table}

Finally, we present upper limits on coalescence rates where the spin of
the components of the binary is taken into account.  Astrophysical
observations of neutron stars indicate that their spins will not be
large enough to have a significant effect on the \ac{BNS} waveform
observed in the \ac{LIGO} band~\cite{ATNF:psrcat,Apostolatos:1994}.
Theoretical considerations limit the magnitude of the spin, $S$, of a
black hole to lie within the range $0 \le S \le G m^{2}/c$.  However,
the astrophysical distribution of black hole spins, and spin
orientations, is not well constrained.  Therefore, we provide a sample
upper limit for spinning systems using a spin magnitude and orientation
distributed uniformly within the allowed values.  This gives upper
limits on the rate of \ac{BBH} and \ac{BHNS} systems of
%
%\begin{eqnarray}
$\mathcal{R}_{90\%,{\rm BBH}} = \SBBHul\,
\textrm{ yr}^{-1}\mathrm{L_{10}}^{-1} \text{ and }
\mathcal{R}_{90\%,{\rm BHNS}} =  \SBHNSul\,
\textrm{ yr}^{-1}\mathrm{L_{10}}^{-1} \, .$
%\end{eqnarray}
%
These rates are about $20\%$ larger than the nonspinning rates.

\paragraph*{Discussion:}

We have searched for gravitational waves from CBCs with total mass
between $2$ and $35\, M_\odot$ in \ac{LIGO} observations between November
14, 2006 and May 18, 2007.  No detection candidates with significance
above that expected due to the background were found in the search. By
combining this search with our previous results, we set a new upper
limit on the CBC rate in the local universe which is approximately a
factor of $3$ lower than that reported in
Ref.~\cite{Collaboration:2009tt}.  This improvement is significant, even
though we searched only two thirds as much data as in
Ref.~\cite{Collaboration:2009tt}.  It is due, in part, to improvements
in detector sensitivity during S5 which increased the horizon distance.
Moreover, the shorter analysis time and improved stationarity of
the data, led to many of the months having a less significant loudest
event than in the previous search.  Both of these effects increased the
luminosity to which the search was sensitive, thereby improving the
upper limit.

Astrophysical estimates for \ac{CBC} rates depend on a number of
assumptions and unknown model parameters, and are still uncertain at
present.  In the simplest models, the coalescence
rates should be proportional to the stellar birth rate in nearby spiral
galaxies, which can be estimated from their blue luminosity
\cite{LIGOS3S4Galaxies}.  The optimistic, upper end of the plausible
rate range for \ac{BNS} is $5 \times 10^{-4} \textrm{ yr}^{-1}
\mathrm{L}_{10}^{-1}$~\cite{Kalogera:2004tn, Kalogera:2004nt} and $6 \times
10^{-5} \textrm{ yr}^{-1} \mathrm{L}_{10}^{-1}$ for \ac{BBH} and \ac{BHNS}
\cite{Oshaughnessy:2008, OShaughnessy:2005}.  
The upper limits reported here are $\sim 1$--$2$ orders of
magnitude above the optimistic expected rates.  With the next run starting in mid 2009, the Enhanced \ac{LIGO} and Virgo
detectors will begin operations with a factor of $\sim 2$ increase in horizon
distance. The total luminosity searched will increase by a factor of $\sim
10$, thereby bringing us close to the optimistic rates.
The most confident \ac{BNS} rate predictions are based on extrapolations
from observed binary pulsars in our Galaxy; these yield realistic
\ac{BNS} rates of $5 \times 10^{-5} \textrm{ yr}^{-1} 
\mathrm{L}_{10}^{-1}$~\cite{Kalogera:2004tn, Kalogera:2004nt}.  Rate
estimates for \ac{BBH} and \ac{BHNS} are less well constrained, but
realistic estimates are $2 \times 10^{-6} \textrm{ yr}^{-1} 
\mathrm{L}_{10}^{-1}$ for \ac{BHNS} \cite{Oshaughnessy:2008} and $4
\times 10^{-7} \textrm{ yr}^{-1} \mathrm{L}_{10}^{-1}$ for \ac{BBH}
\cite{OShaughnessy:2005}.  Thus, the expected rates are $\sim 2$--$3$
orders of magnitude lower than the limits presented in this paper. The
Advanced LIGO and Virgo detectors, currently under construction, will
increase our horizon distance by an order of magnitude or more, allowing us to
measure the rate of CBCs in the Universe.

\paragraph*{Acknowledgements:}
The authors gratefully acknowledge the support of the United States
National Science Foundation for the construction and operation of the
LIGO Laboratory and the Science and Technology Facilities Council of the
United Kingdom, the Max-Planck-Society, and the State of
Niedersachsen/Germany for support of the construction and operation of
the GEO600 detector. The authors also gratefully acknowledge the support
of the research by these agencies and by the Australian Research Council,
the Council of Scientific and Industrial Research of India, the Istituto
Nazionale di Fisica Nucleare of Italy, the Spanish Ministerio de
Educaci\'on y Ciencia, the Conselleria d'Economia, Hisenda i Innovaci\'o of
the Govern de les Illes Balears, the Royal Society, the Scottish Funding 
Council, the Scottish Universities Physics Alliance, The National Aeronautics 
and Space Administration, the Carnegie Trust, the Leverhulme Trust, the David
and Lucile Packard Foundation, the Research Corporation, and the Alfred
P. Sloan Foundation.

%%%%%%%%%%%%%%%%
\bibliography{../bibtex/iulpapers}

\end{document}